\documentclass[journal,comsoc]{IEEEtran}

\usepackage{graphicx}
\usepackage{amsmath}
\usepackage{amssymb}
\usepackage{url}
\usepackage{cite}
\usepackage{xcolor}
\usepackage[table]{xcolor} 
\begin{document}

\title{The Post-Electromagnetic Era: A Vision for Wireless Communication Beyond 6G}

\author{Shumaila  Javaid, and Nasir Saeed,~\IEEEmembership{Senior Member,~IEEE}     
   \thanks{S. Javaid is with the College of Electronics and Information Engineering, Tongji University, Shanghai 201804, and the State Key Laboratory of Autonomous Intelligent Unmanned Systems, Tongji University, Shanghai 201210, China\\
    N. Saeed is with the Department of Electrical and Communication Engineering, College of Engineering, UAE University, Al-Ain 15551, UAE (e-mail: mr.nasir.saeed@ieee.org).}}


\maketitle

\begin{abstract}
Electromagnetic (EM) communication is nearing its physical and thermodynamic limits, where further performance gains through spectrum optimization alone have become increasingly unsustainable. Finite bandwidth, propagation loss at higher frequencies, and the inherent trade-offs between energy and information constrain the scalability of 6G and beyond systems. These limitations drive the search for alternative mechanisms for information transfer beyond conventional EM propagation. This work introduces a state-centric framework for post-6G communication, in which information is conveyed by manipulating physical, biological, and cognitive states rather than EM waves. It identifies ten foundational paradigms that define potential carriers and interaction mechanisms for the post-electromagnetic era and outlines a research roadmap toward self-organizing, cognitively integrated networks. Together, these developments envision a new class of communication systems that are energy-aware, adaptive, and capable of uniting matter, life, and intelligence within a single informational continuum. By establishing the conceptual basis for this transition, the work provides a foundation for future research aimed at realizing communication paradigms that transcend the limitations of spectrum-bound systems.
\end{abstract}

\begin{IEEEkeywords}
Beyond 6G, Post-Electromagnetic Networks, Quantum Communication, Cognitive Continuum, Futuristic Wireless Technologies
\end{IEEEkeywords}

\section{Introduction: From Spectrum to State}
From the advent of First-Generation (1G) cellular systems to ongoing Sixth-Generation (6G) research, wireless communication has evolved through continuous improvements in spectrum utilization, spectral efficiency, and latency reduction. These advancements have been driven by progress in signal processing, transceiver design, and network optimization as 6G systems push the Electromagnetic (EM) paradigm toward its physical and thermodynamic limits. The vision for 7G and beyond points toward communication frameworks that transcend traditional spectrum engineering. These limitations manifest as restricted spatial multiplexing, reduced reliability due to hardware nonidealities, and increasing energy costs for sensing and computation \cite{huang2020holographic,thomas2025survey}. Moreover, in environments such as underwater, subterranean, intra-body, and extraterrestrial settings, absorption and safety limits degrade radio propagation, motivating the development of alternative paradigms beyond EM communication.

\begin{figure}[h!]
    \centering
    \includegraphics[width=0.95\linewidth]{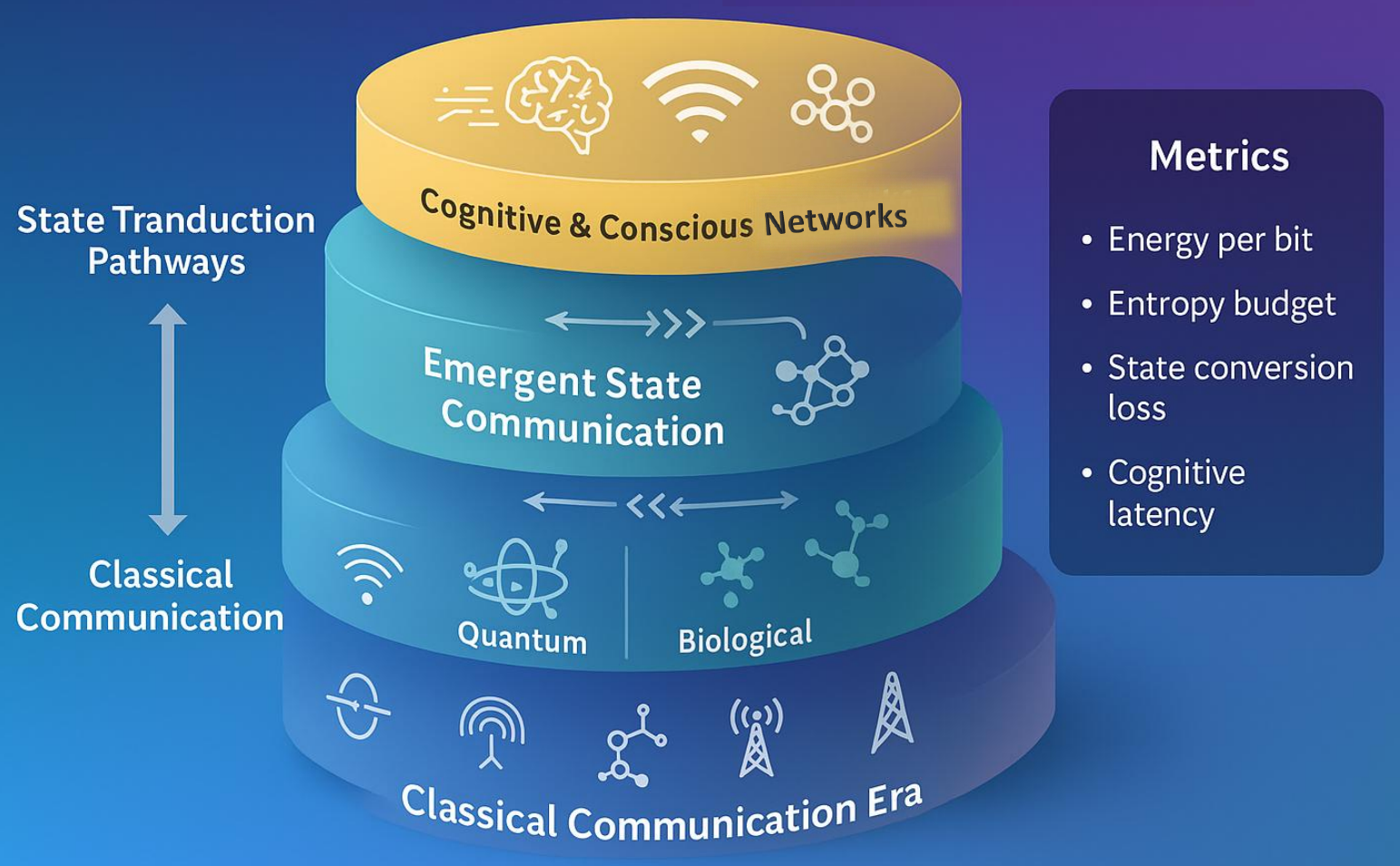}
    \caption{The state continuum for Post-6G communication, showing the evolution from classical EM systems to quantum, biological, and cognitive networks.}
    \label{fig:statecontinuum}
\end{figure}

Progress beyond 6G, therefore, requires a fundamental shift in communication design, from optimizing EM waveforms within a finite spectrum to engineering controlled transitions of physical states across heterogeneous carriers. In the envisioned post-EM paradigm, information is treated as an intrinsic property of matter and energy that can be instantiated, transformed, and conveyed through the medium best suited to a given task or environment. Emerging non-EM carriers, such as spintronic and magnonic channels, phononic waves, and quantum media, extend the communication landscape beyond traditional radio frequencies, where information transfer is no longer confined to oscillations of the EM field but can occur through manipulation of spin, vibration, or quantum state, enabling fundamentally new forms of physical interaction and coupling. EM channels will remain essential \cite{wang2024all}; however, they will coexist within a polymorphic communication fabric, where carrier selection, integration, and transformation become core design principles. This transition is conceptually illustrated in Fig.~\ref{fig:statecontinuum}, which presents the hierarchical evolution from classical EM communication to cognitive and conscious networks through quantum and biological state transduction.
Communication can thus be reframed as state manipulation, unifying signal transport with in situ computation, measurement, and actuation. In this view, networked devices no longer merely radiate and receive signals; they maintain physical states, perform carrier conversions, and dynamically decide whether to communicate, compute locally, or actuate directly based on task-oriented utility. Such coupling necessitates a systems theory that accounts for entropy production and erasure, conversion losses among carriers, and risk-aware attainment of mission objectives. Classical concepts such as capacity and outage must therefore be extended to include convertibility metrics across domains, entropy budgets along end-to-end routes, and information-to-energy ratios that quantify how effectively actions alter the physical state of the environment under energy, safety, and exposure constraints.

Several emerging technologies \cite{zhang2025scalable,potts2025generation} already illustrate this transition. For instance, quantum repeaters and hybrid photon–matter interfaces enable faithful transfer between photonic and material qubits, supporting state teleportation and entanglement distribution over long distances. Spintronic and magnonic interconnects provide energy-efficient short-range links where radiation is inefficient or undesirable. Phononic metamaterials and elastodynamic waveguides guide mechanical energy for robust signaling within structural media. Molecular and ionic communication enables intra-body and lab-on-chip connectivity where radio propagation is limited or regulated. Ambient and backscatter systems further exploit existing fields to achieve ultra-low-power connectivity. When orchestrated by learning-driven controllers that jointly optimize carrier selection, coding, and actuation, these components form adaptive networks capable of responding in real time to environmental variability, energy constraints, and operational risk.

Motivated by these developments, this work introduces a state-centric framework that extends beyond EM communication and emphasizes cross-domain transduction among physical, biological, and cognitive systems. The key contributions of this work are summarized as follows.

\begin{itemize}
    \item The paper presents a conceptual shift from traditional EM communication toward state-based, carrier-agnostic systems. It explores how quantum coherence, biological adaptability, and cognitive awareness can serve as new foundations for information transfer, inspiring research into mechanisms that complement or eventually transcend spectrum-based communication.
    \item The study identifies ten research paradigms, spanning spintronic, quantum-biological, zero-energy, and spacetime communication that collectively outline the emerging architecture of post-EM communication. These pillars form a research roadmap toward integrated, adaptive, and self-organizing networks capable of intelligent environmental interaction.
    \item The work positions the intersection of quantum physics, synthetic biology, and cognitive science as the driving force for next-generation communication. It envisions networks that evolve from data transmission systems into self-maintaining, learning, and co-creative entities, providing a foundation for biologically inspired communication frameworks that are resilient, adaptive, and continuously evolving.
\end{itemize}

\section{The End of the Spectrum Era}
The dominance of EM wave propagation has defined the spectrum era of wireless communication for more than a century. Every generational advancement, from narrowband telephony to emerging 6G technologies, has relied on expanding usable frequencies and improving spectral efficiency through modulation, coding, and spatial multiplexing. However, this dependence on the EM spectrum imposes unavoidable physical, energetic, and environmental constraints. Finite bandwidth, propagation loss, and thermodynamic limits on information processing collectively mark the approaching boundary of spectrum-based progress.
\begin{table}[h!]
\centering
\caption{From Spectrum Optimization to State Engineering}
\renewcommand{\arraystretch}{1.3} 
\setlength{\tabcolsep}{6pt} 

\begin{tabular}{|p{1.5cm}|p{3cm}|p{3.2cm}|}
\hline
\rowcolor{blue!20} 
\textbf{Aspect} & \textbf{Spectrum Era (Classical EM Communication)} & \textbf{Post-EM Paradigm (State-Based Communication)} \\
\hline
\rowcolor{gray!10}
\textbf{Information Carrier} & EM waves (radio, microwave, optical) & Physical states (spin, phonon, quantum, molecular) \\
\hline
\rowcolor{gray!05}
\textbf{Governing Laws} & Maxwell’s equations; Faraday’s and Ampère’s laws & Quantum mechanics, condensed-matter dynamics, field interactions \\
\hline
\rowcolor{gray!10}
\textbf{Encoding Mechanism} & Modulation of electric/magnetic fields & Controlled manipulation of physical or quantum states \\
\hline
\rowcolor{gray!05}
\textbf{Performance Limits} & Bandwidth, SNR, thermodynamic energy cost, aperture loss & Conversion efficiency, coherence lifetime, entropy coupling \\
\hline
\rowcolor{gray!10}
\textbf{Design Focus} & Spectrum allocation and waveform optimization & Cross-domain state transduction and coherence preservation \\
\hline
\rowcolor{gray!05}
\textbf{Example Technologies} & 5G/6G, THz, optical fiber, RIS systems & Spintronic interconnects, phononic links, quantum teleportation \\
\hline
\end{tabular}

\label{tab:spectrum_to_state}
\end{table}
Classical communication, fundamentally governed by Maxwell’s equations, relies on the generation, propagation, and detection of EM fields. Information is encoded in time-varying electric and magnetic fields that radiate into space according to Faraday’s and Ampère’s laws and is ultimately recovered through induced currents at the receiver. This field-based framework underpins all spectrum-based systems, from radio and microwave links to optical communication, and forms the physical foundation of modern wireless design.

Despite its universality, the EM paradigm imposes intrinsic limits on the efficiency with which energy and information can be transmitted. The Shannon–Hartley theorem couples achievable capacity to bandwidth and signal-to-noise ratio, while thermodynamic principles establish a non-zero energy cost for information processing and erasure \cite{yuan2022quantum}. Finite apertures, increasing free-space loss with frequency, and molecular absorption in the Terahertz (THz) range further constrain high-frequency propagation. As these theoretical and practical boundaries are approached, continued performance enhancement through spectrum optimization alone becomes increasingly untenable, motivating exploration of communication principles that transcend the EM domain.

Recent advances in quantum field theory, condensed-matter physics, and nanoscale systems suggest the existence of alternative substrates, such as spin, phonon, or quantum-state interactions that may enable information transfer without classical radiative energy. These developments motivate a transition from spectrum optimization to state engineering, in which communication is realized through controlled transformations of physical states rather than through EM wave propagation. This transition from spectrum-bound communication to state-based information transfer is summarized in Table~\ref{tab:spectrum_to_state}, highlighting the fundamental differences in carriers, governing laws, and design principles between the classical and post-EM paradigms.

\section{Ten Pillars of Post-EM Communication}
\begin{figure*}[h!]
\centering
\includegraphics[width=0.8\textwidth]{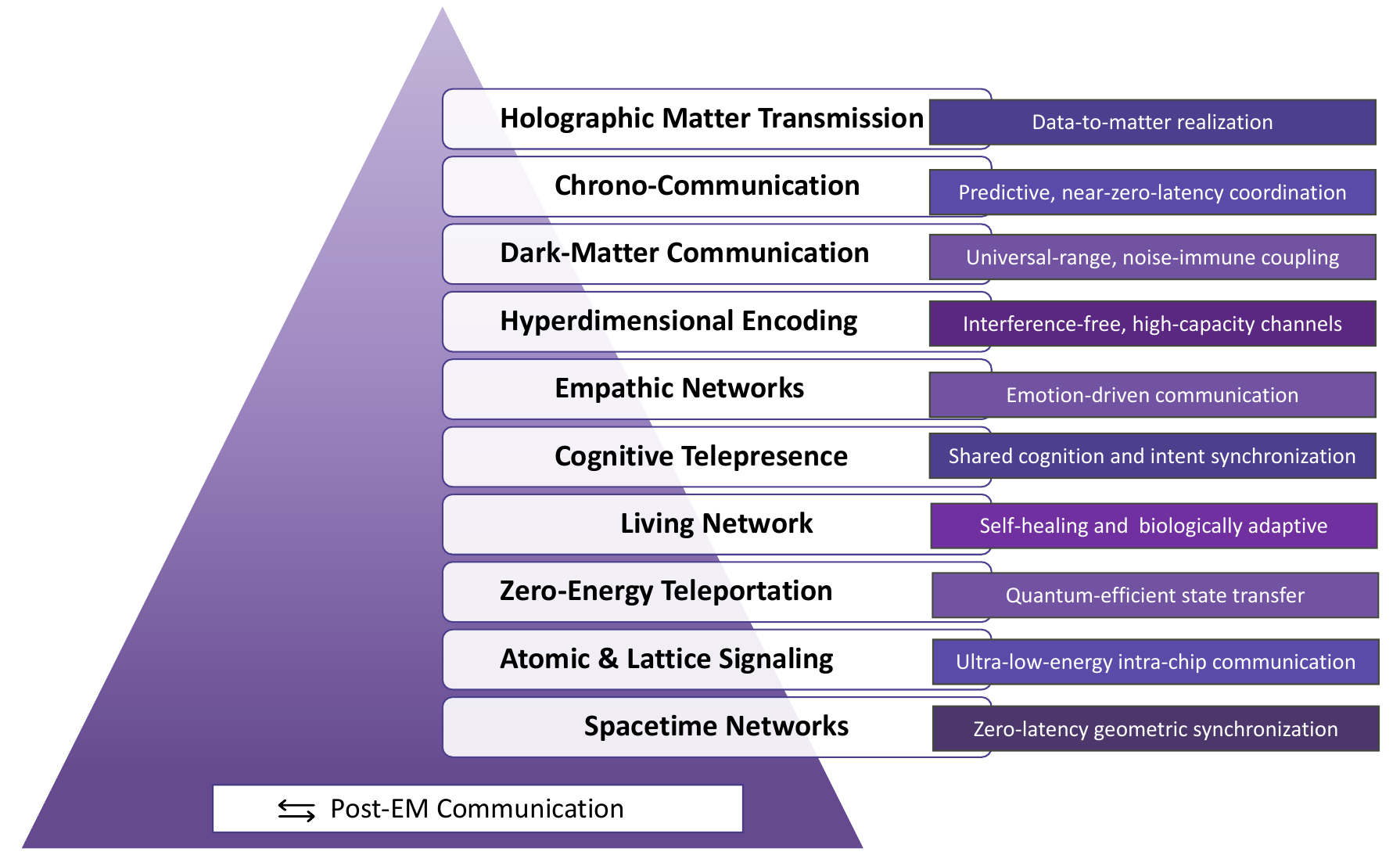}
\caption{Evolution of communication paradigms from physical-state to cognitive and beyond-physical layers.}
\label{fig:pyramid_post_em}
\end{figure*}

This section discusses ten foundational research pillars that define the emerging landscape of post-EM communication. It outlines how information transfer may evolve beyond classical EM propagation toward a unified framework in which communication arises through the direct manipulation of physical, biological, and cognitive states, as illustrated in Figure \ref{fig:pyramid_post_em}.

\subsection{Spacetime Communication Networks}
Spacetime communication envisions the utilization of relativistic or gravitational phenomena to encode, transfer, and synchronize information through the curvature of spacetime rather than through conventional EM fields. In principle, modulating tiny perturbations in spacetime, known as metric fluctuations, could enable information transfer similar to the way EM waves propagate through Maxwellian fields. These perturbations correspond to gravitational waves, whose amplitude is characterized by strain \( h \), typically on the order of \(10^{-21}\) for astrophysical sources. Artificial generation of detectable gravitational waves, however, remains beyond current engineering capabilities because the required energy scales quadratically with both the wave amplitude and frequency \cite{bailes2021gravitational}. Consequently, current research emphasizes indirect or analog approaches that exploit relativistic time transfer and gravitational coupling rather than direct metric modulation.

Although direct modulation of spacetime curvature remains impractical, existing relativistic time-transfer systems demonstrate that spacetime itself can serve as a synchronization and information medium. Spaceborne and optical lattice clocks, together with deep-space timing links, apply general-relativistic corrections to maintain sub-nanosecond synchronization across astronomical distances. Pulsars serve as natural spacetime beacons for navigation and time dissemination, while laboratory-scale optomechanical and acoustic–gravitational analogs provide controlled environments for investigating metric coupling, delay compensation, and phase-coherent time referencing \cite{xia2023entanglement}.

Spacetime communication networks may integrate advances in time–frequency metrology, relativistic geodesy, and quantum-enhanced gravimetry to enable synchronization and information exchange across vast distances without relying solely on EM propagation. The absence of medium-induced dispersion and the deterministic nature of gravitational coupling suggest the potential for latency-invariant coordination, vital for interstellar navigation, deep-space swarm robotics, and precision geospatial referencing. Achieving such systems will require breakthroughs in energy focusing, detection sensitivity, and the suppression of quantum-limited timing noise. Near-term research should therefore prioritize enhancing strain sensitivity below \(10^{-23}\) in compact architectures, quantifying the energy cost per bit for metric perturbations, and maintaining femtosecond-level synchronization stability across astronomical baselines. Relativistic timing networks that combine optical clocks, pulsar references, and entanglement-assisted synchronization offer a practical foundation toward realizing spacetime communication. Ultimately, spacetime communication represents the furthest physical extension of post-EM principles, where information propagates through the geometry of reality itself rather than through any conventional medium.

\subsection{Atomic and Lattice-Level Signaling}
Atomic and lattice level signaling explores the use of quantized spin and vibrational excitations, known as magnons and phonons, as carriers of information at sub-EM scales. Unlike conventional EM waves that radiate energy into free space, spintronic and phononic systems enable localized, low-radiative information transfer within solid-state media. Information can be encoded in the amplitude, phase, or polarization of collective spin precessions and lattice vibrations, enabling compact, energy-efficient, and interference-immune communication channels \cite{wang2024all}

Magnonic communication relies on spin wave propagation in magnetic materials, offering wavelengths orders of magnitude shorter than those of EM signals at comparable frequencies. This enables dense on-chip interconnects, reconfigurable logic, and radiation-free links suited for secure or electromagnetically constrained environments. Challenges include damping, scattering, and mode conversion losses, motivating hybrid transducers that couple magnonic, photonic, and electronic domains with sub-10~dB conversion loss. Similarly, phononic communication employs quantized lattice vibrations to transmit information through mechanical oscillations. Advances in phononic crystals, elastic waveguides, and optomechanical cavities have yielded low-loss, chip-scale acoustic links with high stability and thermal resilience \cite{erdelyi2025design}.

These atomic and lattice-level carriers exchange transmission range for gains in energy efficiency, physical security, and environmental robustness. Their performance is characterized by key metrics such as attenuation length, picojoule-level energy per bit, conversion efficiency, and immunity to crosstalk. With advances in fabrication and material integration, magnonic–phononic communication is expected to reinforce post-EM intra-device and intra-structure networking.

\subsection{Zero-Energy Information Teleportation}
The concept of zero-energy communication must be reframed in physically consistent terms as energy-minimized, entanglement-assisted state transfer. Quantum teleportation transmits information about a quantum state without physically moving the particle. The sender and receiver share a pair of entangled qubits; the sender performs a joint measurement that collapses the state, then transmits two classical bits, allowing the receiver to reconstruct the original state locally. While teleportation itself requires no propagation of the encoded signal, generating and maintaining entanglement, as well as operating the classical side channel, consume energy bounded by thermodynamic principles such as Landauer’s limit, which defines the minimum energy required to process or erase one bit of information \cite{hu2023progress}.  
Recent advances in quantum repeaters, hybrid photon–matter interfaces, and solid-state quantum memories have made long-distance entanglement distribution increasingly feasible. Networks employing superconducting qubits, trapped ions, and nitrogen–vacancy centers now demonstrate entanglement fidelities above 0.9 across multi-node configurations \cite{mantri2025comparing}. Such architectures provide the foundation for near-zero-energy communication, in which information transfer occurs via quantum-state reconstruction rather than direct radiative exchange.  
From an engineering perspective, achieving high-efficiency quantum teleportation requires optimizing three interdependent parameters, including the entanglement rate (pairs generated per second), teleportation fidelity, and classical-bit overhead. The total energy per teleported qubit accounts for both entanglement distribution and classical signaling. Ongoing work focuses on minimizing this cost through photonic integration, low-loss quantum memory materials, and on-demand entanglement generation, bringing communication closer to the thermodynamic limit of reversibility. Energy-minimized teleportation thus exemplifies the broader post-EM transition, where information is encoded, transformed, and reconstructed through the manipulation of physical states rather than EM carriers.

\subsection{Living Networks and Biological Communication}
Living networks represent a paradigm in which biological systems themselves function as dynamic communication media. Bio-hybrid substrates, such as neural organoids, engineered tissues, and molecular or ionic pathways, enable adaptive, self-healing links in environments where EM propagation is inefficient or biologically constrained. In such systems, information can be carried by ionic currents, chemical gradients, or neuron-like spiking patterns, allowing for energy-efficient and intrinsically integrated sensing, processing, and actuation within physiological or microfluidic domains \cite{pappalardo2025synthetic}.

Biological communication mechanisms are inherently robust to structural perturbations and exhibit emergent repair and adaptation capabilities unavailable in conventional electronic systems. Recent studies have demonstrated living neuronal cultures capable of learning and responding to external stimuli, forming the basis for in vitro “wetware” computing and signal transfer. Similarly, ionic and molecular signaling pathways enable short-range information transmission through diffusion and binding dynamics, supporting intra-body and lab-on-chip connectivity where RF-based communication is attenuated or regulated. These biological substrates offer unique properties such as self-repair, growth-based reconfiguration, and biocompatibility, which can be leveraged for implantable sensors, prosthetic interfaces, and distributed biosensing systems.  
However, realizing reliable biological communication faces several challenges, including biocompatibility, long-term stability, variability among living cells, and biosafety concerns. Critical factors in system design include maintaining low bit-error rates in physiological media, ensuring long-term link stability, minimizing immune response, and achieving reliable closed-loop actuation. Hybrid bioelectronic systems that integrate organic tissue interfaces with micro and nanoelectronic control are emerging as a promising approach for establishing controllable, interpretable communication within living substrates. Accordingly, living networks and biological communication extend the post-EM paradigm by transforming biological matter into an active carrier of information. They exemplify how the convergence of synthetic biology, neuromorphic engineering, and molecular communication can establish energy-efficient, self-sustaining channels that merge computation, sensing, and actuation within the fabric of life itself.

\subsection{Consciousness and Cognitive Telepresence}
Cognitive telepresence envisions the direct transmission and synchronization of perceptual and intentional states across spatially separated entities. It extends beyond conventional Brain–Computer Interfaces (BCIs) toward bidirectional neural links capable of sharing high-level cognitive representations such as intent, emotion, and learned behavior between humans and artificial agents. In principle, such links could enable task-level collaboration and cooperative control without linguistic intermediaries, reducing communication latency to the scale of neural processing.

Recent advances in noninvasive neural sensing, real-time decoding, and closed-loop stimulation have accelerated BCI development by integrating machine learning with neural electrophysiology. Experiments have demonstrated brain-to-brain transfer of motor intent and visual stimuli in both rodents and humans using Electroencephalography (EEG) and transcranial focused ultrasound \cite{kosnoff2024transcranial}. Advances in optogenetics, high-density electrode arrays, and nanoscale neural dust sensors further enhance spatial and temporal resolution, enabling reliable acquisition and delivery of multimodal neural signals.

A cognitive telepresence network integrates sensing, inference, and actuation layers within a closed-loop neural communication architecture that preserves agency, privacy, and cognitive integrity through secure encoding and adaptive feedback. In this context, consciousness-oriented communication marks a frontier in post-EM networking, where information is instantiated through the direct coupling of cognitive states rather than conveyed through symbolic transmission.

\subsection{Emotionally Adaptive and Empathic Networks}
Emotionally adaptive networks extend human-centered communication by integrating affective and physiological sensing into network control and resource allocation. These systems interpret emotional, cognitive, and physiological states in real time to adjust parameters such as bandwidth, latency, and data priority according to user well-being and contextual demands. The resulting \emph{Internet of Emotions} envisions infrastructure responsive to both traffic conditions and collective human states.
The principal advantage of emotionally adaptive networks lies in their ability to align communication quality with human affective states, thereby reducing cognitive load and improving perceived quality of experience. Unlike static network architectures, these systems continuously adapt to user context, ensuring both technical efficiency and emotional well-being, an essential characteristic of next-generation human-centric communication infrastructure. Advances in wearable biosensors, multimodal affective computing, and emotion-aware signal processing form the foundation of empathic connectivity. Noninvasive sensing methods such as Photoplethysmography (PPG), Electrodermal Activity (EDA), facial electromyography, and EEG enable continuous inference of stress, engagement, and cognitive load. Machine learning models, particularly transformer-based multimodal fusion networks, have achieved high accuracy in decoding emotional valence and arousal across diverse contexts \cite{ li2024real}. As a result, these developments position emotionally adaptive and empathic networks as a new interface between humans and communication infrastructure. By embedding emotional intelligence into the network fabric, such systems bridge biological affect and digital communication, aligning network performance with collective psychological and physiological equilibrium.

\subsection{Hyperdimensional Information Embedding}
Hyperdimensional information embedding extends communication capacity beyond conventional spectral expansion by encoding data into high-dimensional state spaces. Instead of relying solely on frequency, time, and amplitude, these systems employ additional orthogonal physical modes such as photonic Orbital Angular Momentum (OAM), spin-wave distributions, and topological or mechanical resonances to create multiple independent channels within a shared carrier. Each dimension represents a distinct degree of freedom through which information can be modulated, transmitted, and decoded, thereby multiplying effective capacity without increasing spectral bandwidth.

Recent advances in optical, spintronic, and mechanical platforms have demonstrated high-dimensional mode manipulation for data transmission. Photonic OAM multiplexing enables encoding across multiple helical phase states, achieving terabit-per-second throughputs in free-space and fiber links \cite{wang2022orbital}. Magnonic and phononic devices support dense on-chip and near-field communication through multimodal excitation, while topological and mechanical metamaterials provide robust mode isolation and backscattering immunity for stable propagation. The performance of hyperdimensional systems depends on the number of usable orthogonal modes, inter-mode crosstalk, insertion loss, and conversion fidelity, with energy cost per encoded mode defining scalability. Machine-learning-based demultiplexing and adaptive mode control further enhance real-time compensation for distortion and turbulence in photonic and acoustic domains. These developments establish hyperdimensional information embedding as a central pillar of post-EM communication, enabling interference-resilient multidimensional data exchange across physical and virtual environments.

\subsection{Dark Matter Communication}
Dark-matter communication represents a speculative yet conceptually intriguing frontier within post-EM research. If Weakly Interacting Massive Particles (WIMPs) or other non-luminous field components beyond the Standard Model could be measurably coupled to engineered materials, they might enable ultra-low-interference and long-range communication channels. Such media would, in principle, be immune to conventional EM noise, offering universal-range connectivity across extreme environments such as interstellar or subterranean domains.  

At present, no empirical framework exists for deliberate modulation or detection of dark-matter-based signals. Research is limited to theoretical coupling models, astrophysical detection experiments, and quantum sensing protocols that probe ultra-weak field interactions. Practical evaluation of these phenomena relies on constraints imposed by the detection cross-section, the thermal noise floor, and the achievable information rate under energy and sensitivity limitations. Although still in its early stages, exploration of dark-matter communication underscores the ambition of post-EM paradigms to identify and exploit previously inaccessible physical substrates for information transfer. Its study reveals the fundamental limits of coupling, detection, and energy conversion across exotic media, defining the boundaries of physically realizable communication.

\subsection{Chrono-Communication}
Chrono-communication explores the boundary between predictive control and physical causality in pursuit of effectively zero-latency communication. Rather than violating temporal order, it leverages prediction, synchronization, and shared state estimation to reduce perceived delay. Techniques such as pre-shared randomness, predictive coding, and entanglement-assisted time transfer allow distributed agents to act on anticipated information before explicit arrival, while maintaining causal consistency through classical feedback and model correction. 

From an implementation perspective, chrono-communication integrates precise clock synchronization, low-jitter timing distribution, and adaptive learning models to enable predictive coordination across distributed systems. Performance can be evaluated through metrics such as prediction horizon and error, effective latency reduction, and stability under model drift or non-stationary conditions. These capabilities enhance coordinated autonomy, teleoperation, and immersive virtual interaction, where timing precision directly affects control stability and user experience. By reframing communication as anticipatory coordination rather than post-event exchange, chrono-communication shows how predictive intelligence and temporal discipline can approximate zero-latency operation within the limits of relativity, aligning temporal coherence with the broader post-EM objective of state-level synchronization.

\subsection{Holographic Matter Transmission}
Holographic matter transmission generalizes communication from information exchange to physical-state reconstruction. Rather than literal matter teleportation, this concept focuses on transmitting compressed atomic or mesoscopic descriptors of an object (e.g., geometry, composition, and dynamic properties) and reproducing them remotely using programmable materials, self-assembling nanostructures, or additive manufacturing integrated with digital twin systems.  
Recent advances in nanoscale 3D printing, metamaterial reconfiguration, and molecular self-assembly are opening early pathways toward state-to-matter reproduction. Embedding material blueprints directly into data streams extends communication from digital representation to tangible realization. Key performance indicators for such systems include structural fidelity, quantified through root-mean-square reconstruction error, fabrication time, energy required per reconstructed entity, and the security of transmitted blueprints. Holographic matter transmission thereby unites the informational and physical domains, transforming communication into a bidirectional process of encoding, transporting, and realizing material structure. It represents the ultimate convergence of data, energy, and material transformation envisioned within the post-EM communication paradigm.

\section{Convergence of Physics, Biology, and Consciousness}
\begin{figure}[t]
    \centering
    \includegraphics[width=0.8\linewidth]{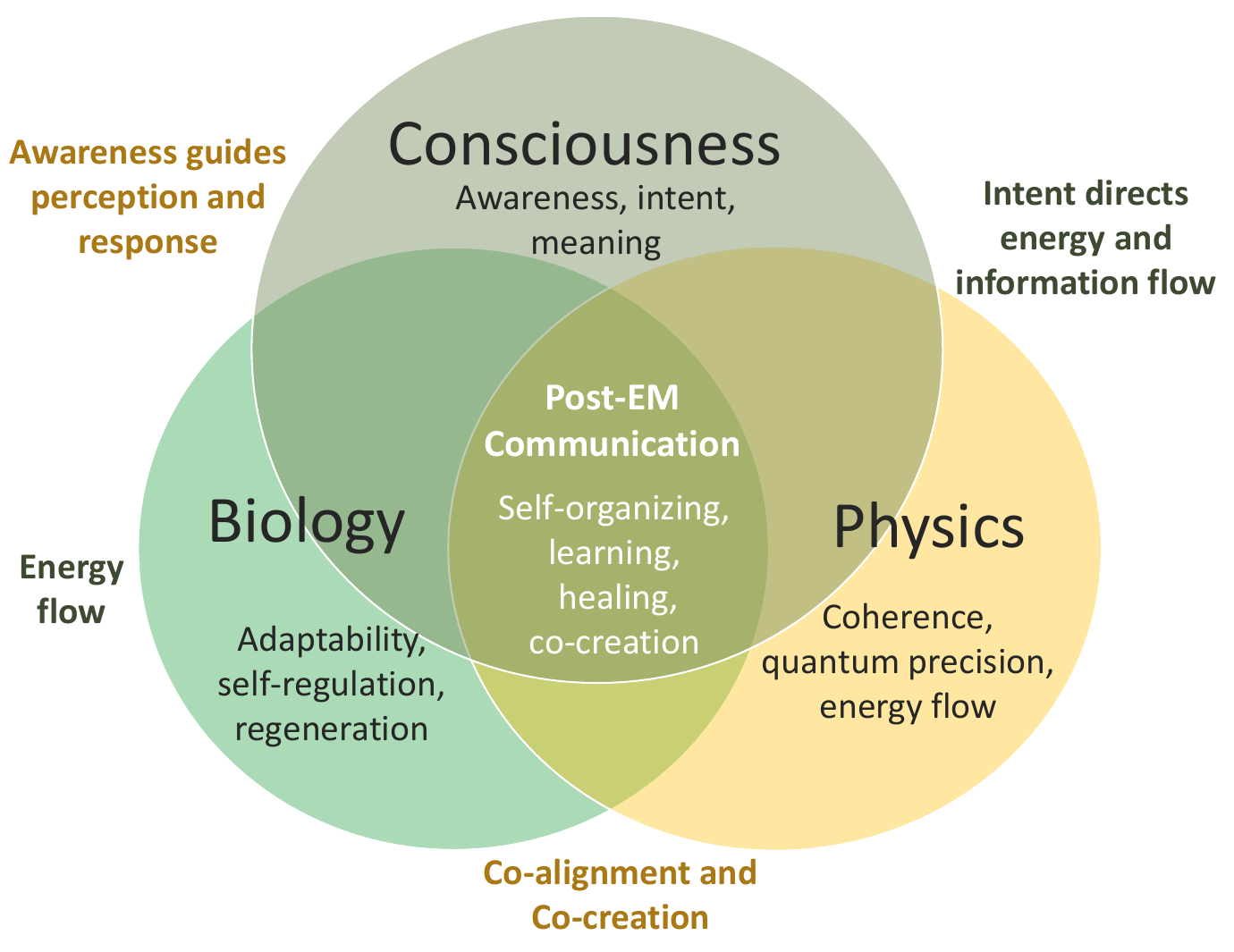}
    \caption{Conceptual illustration of post-EM communication arising from the convergence of physics, biology, and consciousness. Their integration, combining coherence, adaptability, and awareness, enables self-organizing, learning, and co-creative communication networks.}
    \label{fig:convergence_framework}
\end{figure}

This section explores how the convergence of quantum physics, synthetic biology, and cognitive science is redefining the foundations of communication in the post-EM era. It aims to show that communication is no longer confined to the exchange of EM signals but extends into the coordinated evolution of physical, biological, and cognitive states. By integrating quantum coherence, biological adaptability, and cognitive awareness within a unified framework, this convergence transforms communication systems from engineered infrastructures into self-evolving, living networks capable of learning, healing, and co-creating meaning across the continuum of matter, life, and mind, as illustrated in Fig.~\ref{fig:convergence_framework}.

\subsection{Quantum–Biological Interfaces}
Recent research in quantum biology shows that nature already exploits quantum phenomena for efficient energy and information transfer. Coherence lifetimes in photosynthetic complexes, for instance, have been experimentally observed in the range of 200–600 femtoseconds, allowing excitons to explore multiple energy pathways simultaneously and minimizing loss. Similarly, spin-dependent reactions in avian magnetoreception and tunneling in enzymatic catalysis suggest that living systems sustain quantum correlations under physiological conditions.  

Synthetic biology seeks to replicate and engineer these effects by embedding quantum sensors and spintronic elements into biological scaffolds. Quantum–biological hybrids could thus serve as ultra-sensitive, self-repairing communication nodes capable of both sensing and responding to their environment. These systems merge the precision of quantum physics with the adaptability of living matter, forming the substrate for biophysically coherent information exchange. Thus, they establish a bridge between quantum-level precision and biological intelligence, pointing toward communication systems that operate through coherent state interactions rather than EM signaling.

\subsection{Cognitive and Neural Coupling}
At the cognitive frontier, advances in BCIs are creating direct bidirectional links between neural activity and machine intelligence. Modern noninvasive systems using EEG and Functional Near-Infrared Spectroscopy (FNIRS) achieve transfer rates of 150–250 bits per minute for intent decoding, with emerging multimodal neural fusion methods pushing these limits further. Distributed cortical meshes, containing thousands of recording and stimulation nodes, are enabling real-time feedback loops that integrate perception, computation, and action. When combined with neuromorphic processors, these systems allow shared learning between biological and artificial agents, effectively merging cognition and computation. Communication, in this context, becomes a process of state alignment, the continuous synchronization of neural, computational, and environmental patterns. Together, they point toward a future in which communication emerges from the seamless co-evolution of human cognition and machine intelligence, enabling networks that adapt and learn in unison with their users.

\subsection{Self-Evolving Communication Ecosystems}

As quantum and biological systems gain cognitive agency, communication networks evolve from static architectures into adaptive ecosystems. Nodes no longer act as passive transmitters but become intelligent agents that learn from interactions, self-organize, and restore functionality after disruption. Experimental prototypes already demonstrate this transition. Living neural circuits integrated with microelectronic controllers exhibit autonomous adaptation, while synthetic cells are being programmed to exchange molecular messages and collectively regulate behavior.  

Such systems redefine the criteria for communication performance. Instead of focusing solely on throughput or latency, the emphasis now lies on resilience, adaptability, and informational consistency. These living networks function as self-maintaining entities whose internal organization evolves to preserve coherence, recover from disturbances, and sustain purposeful operation across multiple scales. As communication systems begin to integrate physical coherence, biological adaptability, and cognitive awareness, they move beyond technical functionality into domains traditionally associated with life and intelligence. This blurring of boundaries between engineered and living systems inevitably raises questions about agency, identity, and moral responsibility in networks that can sense, decide, and evolve autonomously.

\section{Ethical and Existential Implications}
The emergence of post-EM communication extends engineering into the domain of existence itself. When physical, biological, and cognitive substrates merge, communication becomes inseparable from creation and transformation. Such deep integration raises profound ethical and philosophical questions. Who owns transmitted consciousness? How can authenticity be verified when information can alter or instantiate matter? What constitutes consent when interaction occurs between biological and synthetic minds?
Traditional frameworks for data governance and cybersecurity are inadequate for living, self-organizing networks that exhibit awareness or adaptive intent. Ethical design must therefore evolve from compliance-based regulation to continuous, context-sensitive stewardship. Principles such as transparency, autonomy preservation, and accountability must be encoded in software as well as in the physics and biology of the communication substrate itself.

At a societal level, the capacity to transmit thought, emotion, or embodied experience challenges the foundations of privacy, individuality, and even mortality. The distinction between sender and receiver, or human and machine, becomes fluid, requiring new forms of legal identity and relational ethics. Education, policy, and technology must converge to ensure that such capabilities serve collective well-being rather than exploitation or control. Accordingly, the post-EM era compels humanity to reconsider communication not merely as an exchange of information but as an act of co-creation, shaping the continuum of matter, life, and consciousness. As communication extends across physical, biological, and cognitive domains, its stewardship must reflect both scientific insight and ethical wisdom, ensuring that increasingly autonomous networks remain aligned with human values and planetary sustainability.

\section{Research Roadmap to the Cognitive Continuum}
The evolution toward post-EM communication is expected to progress through multiple generational milestones. Each phase extends communication capabilities from intelligent spectrum control toward the integration of quantum, biological, and cognitive processes, eventually forming a unified cognitive continuum. The main milestones are outlined below and summarized in Table I.

\begin{itemize}
    \item \textbf{Phase I (6G–7G, 2025–2035):} 
    Development of AI-defined spectrum management, THz photonic communication, and Reconfigurable Intelligent Surfaces (RIS 3.0). These advances will enable adaptive control of EM environments and form the computational basis for context-aware networking.
    
    \item \textbf{Phase II (7.5G, 2035–2045):}
    Convergence between quantum physics and biology will give rise to quantum–biological transceivers and photonic neural chips, allowing communication systems to sense, learn, and self-regulate at the physical layer.
    
    \item \textbf{Phase III (8G, 2045–2055):}
    Exploration of zero-energy vacuum signaling and dark-matter modulation will extend communication beyond classical EM propagation, focusing on non-radiative energy exchange and information transfer across exotic physical media.
    
    \item \textbf{Phase IV (9G and Beyond, 2055–2070):}
    Chrono-communication and hyperdimensional embedding will redefine time and space as active information carriers, leading to predictive, self-evolving networks that operate across the continuum of matter, life, and consciousness.
\end{itemize}

\begin{table}[h!]
\centering
\caption{Milestones Toward Post-EM Communication}
\label{tab:roadmap}
\renewcommand{\arraystretch}{1.3} 
\setlength{\tabcolsep}{8pt} 

\definecolor{headerblue}{RGB}{210,225,250}
\definecolor{rowgray}{gray}{0.95}

\begin{tabular}{|p{2cm}|p{2cm}|p{3cm}|}
\hline
\rowcolor{headerblue}
\textbf{Phase} & \textbf{Timeline} & \textbf{Key Breakthroughs} \\
\hline
\rowcolor{rowgray}
6G--7G & 2025--2035 & AI-defined spectrum, THz photonics, RIS 3.0 \\
\hline
7.5G & 2035--2045 & Quantum-biological transceivers, photonic neural chips \\
\hline
\rowcolor{rowgray}
8G & 2045--2055 & Zero-energy vacuum signaling, dark-matter modulation \\
\hline
9G & 2055--2070 & Chrono-communication, hyperdimensional channels \\
\hline
\end{tabular}

\end{table}

The roadmap illustrates that the journey toward the cognitive continuum is a linear extension of wireless technologies as well as a paradigm shift, from optimizing EM spectra to engineering communication as an expression of intelligent physical and biological state evolution.

\section{Conclusion: The Cognitive Continuum}

This work presents a forward-looking framework that redefines communication beyond the EM paradigm. It introduces the concept of state-centric communication, in which information is conveyed through the controlled manipulation and transformation of physical, biological, and cognitive states rather than through conventional EM waves. By framing this transition, the paper establishes a unified foundation for exploring post-EM information transfer across quantum, biological, and cognitive domains. Ten foundational paradigms are outlined to illustrate the diverse pathways through which post-EM communication may evolve, ranging from spacetime signaling and atomic-scale modulation to cognitive telepresence and holographic matter transmission. A research roadmap is also introduced to guide progress across future network generations, highlighting how communication may develop from AI-defined spectrum control in 6G and 7G to fully cognitive, self-evolving networks in 9G and beyond.

The vision shows that the future of communication lies in redefining the very nature of connectivity itself. This transformation envisions networks that are adaptive, intelligent, and ethically aligned, capable of integrating matter, life, and awareness into a coherent informational continuum. The post-EM era therefore, represents both a scientific and a philosophical evolution in how information is perceived, processed, and shared across the universe.

\bibliographystyle{IEEEtran}
\bibliography{Ref}
\end{document}